\documentstyle[twocolumn,aps,epsf]{revtex}

\def\myfigure#1#2{{\leftskip=0.000753\textwidth \rightskip\leftskip\small
\begin{figure}\baselineskip=14pt plus 2pt minus 1pt
\centerline{#1}\nobreak\smallskip\nobreak #2\end{figure}}}
\global\firstfigfalse

\draft	

\begin{document}
\title{Shape Deformation driven Structural Transitions in Quantum Hall Skyrmions}

\author{Madan Rao$^{1}$\cite{MAD}, Surajit Sengupta$^2$\cite{SUR} and
R. Shankar$^{1}$\cite{SHANKS}}

\address{$^1$Institute of Mathematical Sciences, Taramani, 
Chennai (Madras) 600113,
India\\ $^2$Material Science Division, Indira Gandhi Centre for Atomic
Research, Kalpakkam 603102, India}

\date{\today}

\maketitle

\begin{abstract}

The Quantum Hall ground state away from $\nu = 1$ can be described by a
collection of interacting skyrmions.  We show within the context of 
a nonlinear sigma model, that the classical ground state away from
$\nu = 1$ is a skyrmion crystal with a generalized N\'eel order. We show
that as a function of filling $\nu$, the skyrmion crystal undergoes a
triangle $\to$ square $\to$ triangle transition at zero temperature. We
argue that this structural transition, driven by a change in the shape of 
the individual skyrmions, is stable to thermal and quantum fluctuations 
and may be probed experimentally. 

\end{abstract}

\bigskip


Following a recent suggestion \cite{dhlee,sondhi,fertig1}, there has been growing
experimental evidence \cite{barret,aifer,schmeller} that the charged quasiparticle
excitations about the $\nu = 1$ quantum Hall state in GaAs heterostructures are
extended objects called skyrmions with spin significantly greater than $1 /2$. Recent
OPNMR \cite{barret} and optical magneto-absorption \cite{aifer} experiments observe a
sharp fall in the spin polarization as the filling factor is changed from $\nu = 1$,
consistent with the existence of skyrmions. 

What is the ground state of a system of interacting skyrmions in two dimensions ?
Specific heat measurements \cite{bayot} on GaAs heterojunctions at $\nu = 0.77$, show
a sharp peak at a temperature $T_c \approx 30$mK.  It has been suggested \cite{bayot}
that this anomaly may be associated with the freezing of skyrmions into a crystal
lattice. Previous theoretical work \cite{fertig2,kogan} have analysed the crystalline
state of skyrmions. Using a mean-field analysis of electrons confined to the lowest
Landau level, Fertig {\it et. al.} \cite{fertig2} claim that at $T=0$, the skyrmions
form a square lattice with a N\'eel orientation ordering.  On the other hand, by a
mapping onto an effective nonlinear sigma model, Green {\it et. al.} \cite{kogan}
conclude that the skyrmions form a triangular lattice with a generalized N\'eel
ordering. 

In this paper, we calculate the $T=0$ classical phase diagram of a system of
interacting skyrmions using an effective classical O(3) nonlinear sigma model (NLSM).
Within a variational scheme, we show that at values of $\nu$ sufficiently away from
$1$, the ground state is a triangular lattice of skyrmions with a three sublattice
generalized N\'eel orientation order. As we approach $\nu=1$, we find a structural
transition to the square lattice with N\'eel orientation order. This structural
transition is accompanied by a change in the shape of the individual skyrmions and a
jump in the spin polarization. In the dilute limit, $\nu \to 1$, we find a reentrant
triangular phase which is different from the previously encountered triangular phase!
We also provide arguments for the stability of these phases to thermal and quantum
fluctuations. 

The low energy excitations about the ferromagnetic $\nu=1$ ground state in GaAs
hetrostructures are described by an NLSM \cite{dhlee,sondhi}, in terms of a unit
vector field that represents the local spin polarization. Neutral excitations
correspond to spin waves, while charged excitations correspond to defect
configurations (skyrmions) with topological charge $Q$ (the electronic charge density
is equal to the topological charge density $ 4\pi q({\bf x})={\bf n}\cdot(\partial_x
{\bf n}\times \partial_y {\bf n})$ in the long wavelength limit\cite{dhlee}). The
topological charge $Q = \int q({\bf x})\,d^2x$ is always an integer and counts the
number of times the spin configuration ${\bf n}({\bf x})$ wraps around the unit
sphere. 

For convenience we work in a notation where the unit vector field ${\bf n} (x, y)$ is
replaced by a complex field $w (z \equiv x+iy, \bar z \equiv x-iy)$, obtained by the
stereographic projection of the unit sphere onto the complex plane. Thus
$w=cot({\theta \over 2})~e^{i\phi}$, where $\theta$ and $\phi$ are the polar angles
of the unit vector ${\bf n}$. For static, classical spin configurations, the energy
functional \cite{sondhi} may be written as,
\begin{equation}
E\left[\bar w, w\right] = E_0\left[\bar w, w\right] + E_{Z}\left[\bar 
w, w\right] + E_{coul}\left[\bar w, w\right] \,\,.
\label{eq:etot}
\end{equation}
The first term on the right 
\begin{equation}
E_0\left[\bar w, w\right]= \gamma \int_{z,\bar z}
         {1 \over (1+\bar w w)^2}
	(\partial_{\bar z}{\bar w}\partial_zw +\partial_z{\bar w}\partial_
        {\bar z} w )
\label{eq:e0}
\end{equation}
describes spin exchange. The Zeeman coupling of the spins to the external magnetic
field $B$ is given by
\begin{equation}
E_{Z}\left[\bar w, w\right] = {g^{*} \over 2\pi}\int_{\bar z,z}
                   {\bar w w \over (1+\bar w w)} 
\label{eq:ez}
\end{equation}
while the charged quasiparticles interact via a long-range coulomb interaction,
\begin{equation}
E_{coul}\left[\bar w, w\right] = {e^{*} \over 2}\int_{\bar z,z}\int_{\bar 
                  z^\prime,z^\prime}
		{q(\bar z,z)q(\bar z^\prime,z^\prime) \over |z-z^\prime|}
                \,\,.
\label{eq:ecoul}
\end{equation}
The energy functional Eq.\ (\ref{eq:etot}) has been expressed in dimensionless
variables, with all energies in units of the cyclotron energy $\hbar
\omega_c,~\omega_c=eB/m^*c$ ($m^*$ is the electron band mass) and all lengths in
units of the magnetic length $l_c, ~2\pi l_c^2=hc/eB$. The parameters in the energy
functional are $\gamma=e^*/(16\sqrt{2\pi})$, $g^{*}=g\mu_BB/ \hbar \omega_c$ and
$e^{*}=(e^2 / K l_c)(1/\hbar \omega_c)$ ($K$ is the dielectric constant of GaAs). 

In the absence of Coulomb and Zeeman interactions, any (anti)meromorphic function
$w(z)$ is a solution of the resulting Euler Lagrange equations \cite{polya}.  The
topological charge is simply $Q = \sum_i n_i$, where $i$ runs over the poles of $w$
and $n_i$ is the degree of the $i^{th}$ pole. The one skyrmion solution given by
\begin{equation}
w(\bar z,z)= {\lambda e^{i\Omega} \over z-\xi} \,\,,
\label{eq:1sky}
\end{equation}
clearly has a $Q = 1$. The spin and charge distributions are centred at $\xi$ and
fall off as power laws with a scale set by $\lambda$. The $XY$ component of the spin
at ${\bf x}$ is oriented at an angle $\Omega$ to the position vector ${\bf x}$. The
$Z$ component of the spin $S_Z$, however, diverges logarithmically. 

In the presence of the Zeeman and Coulomb interactions, Eq.\ (\ref{eq:1sky}) is no
longer the minimum energy solution. These terms destroy scale invariance, and
generate a `size' for the optimal skyrmion, leading to a finite $S_Z$.  It is natural
to try the following variational form for the single skyrmion solution which
minimises Eq.\ (\ref{eq:etot}),
\begin{equation}
w(\bar z,z)= {\lambda e^{i\Omega} \over z-\xi}e^{-\kappa \vert z-\xi\vert}\,\,.
\label{eq:1ksky}
\end{equation}
With $\lambda$ and $\kappa$ as variational parameters, we have checked that the
lowest energy single skyrmion solution has a $\kappa > 0$. The topological charge
remains $1$, since the configuration in Eq.\ (\ref{eq:1ksky}) can be smoothly
deformed to the configuration in Eq.\ (\ref{eq:1sky}). 

A system of $N$ identical $Q=1$ skyrmions centred at $\{ \xi_I \}$ with orientations
$\{ \Omega_I \}$, can now be parametrized by
\begin{equation}
w(\bar z,z) = \sum_{I=0}^N{\lambda e^{i\Omega_I} \over z-\xi_I}
e^{-\kappa \vert z-\xi_I\vert}\,\,.
\label{eq:nsky}
\end{equation}
The topogical charge $Q = N$, however we observe that even for $\kappa=0$, the spin
$S_Z$ is finite. This illustrates the fact that the $Z$ component of the total spin
of overlapping skyrmions is not the sum of the individual spins.  Our crystalline
ansatz corresponds to placing the $\{ \xi_I \}$ on a triangular or a square lattice.
We have studied both the ferro-oriented ($\Omega_I=0, \forall I$) and generalized
N\'eel oriented (described below) configurations. We find that the ferro-oriented
configurations always have a higher energy and so shall ignore them for the rest of
this paper. Since the square lattice is bipartite, the N\'eel configuration is
characterised by $\Omega_I = 0$ for the $A$ sublattice and $\Omega_I=\pi$ for the $B$
sublattice. Likewise, for the tripartite triangular lattice, the generalized N\'eel
ordering is obtained by assigning $\Omega_I = 0, 2\pi/3, 4\pi/3$ to the $A, B$ and
$C$ sublattices respectively. These orientation assignments emerge naturally in the
$\kappa =0$ limit of Eq.\ (\ref{eq:nsky}), when it reduces to an elliptic function.
The sum of the residues of an elliptic function in the unit cell should equal zero.
In the present context this is precisely the generalized N\'eel condition that the
sum of the orientations of all the skyrmions in a unit cell is zero (mod $2\pi$). 

To find the classical ground states, we numerically compute the energy (accurate to
$1$ part in $10^{6}$) using our crystalline ansatz and minimize with respect to the
variational parameters $\kappa$ and $\lambda$ for a given lattice and orientational
ordering. We have chosen typical values of the carrier concentration ($n = 1.5 \times
10^{11}$ cm$^{-2}$) and magnetic fields ($B = 6 - 10$ Tesla) and have varied $\nu$ by
tilting the magnetic field with respect to the normal keeping its magnitude fixed.
The lattice spacing $a$ for a given lattice type is fixed by the value of $\nu$
(e.g., for a square lattice, $a \equiv \sqrt{2\pi/\vert1-\nu\vert}$). 
\myfigure{\epsfysize2.5in\epsfbox{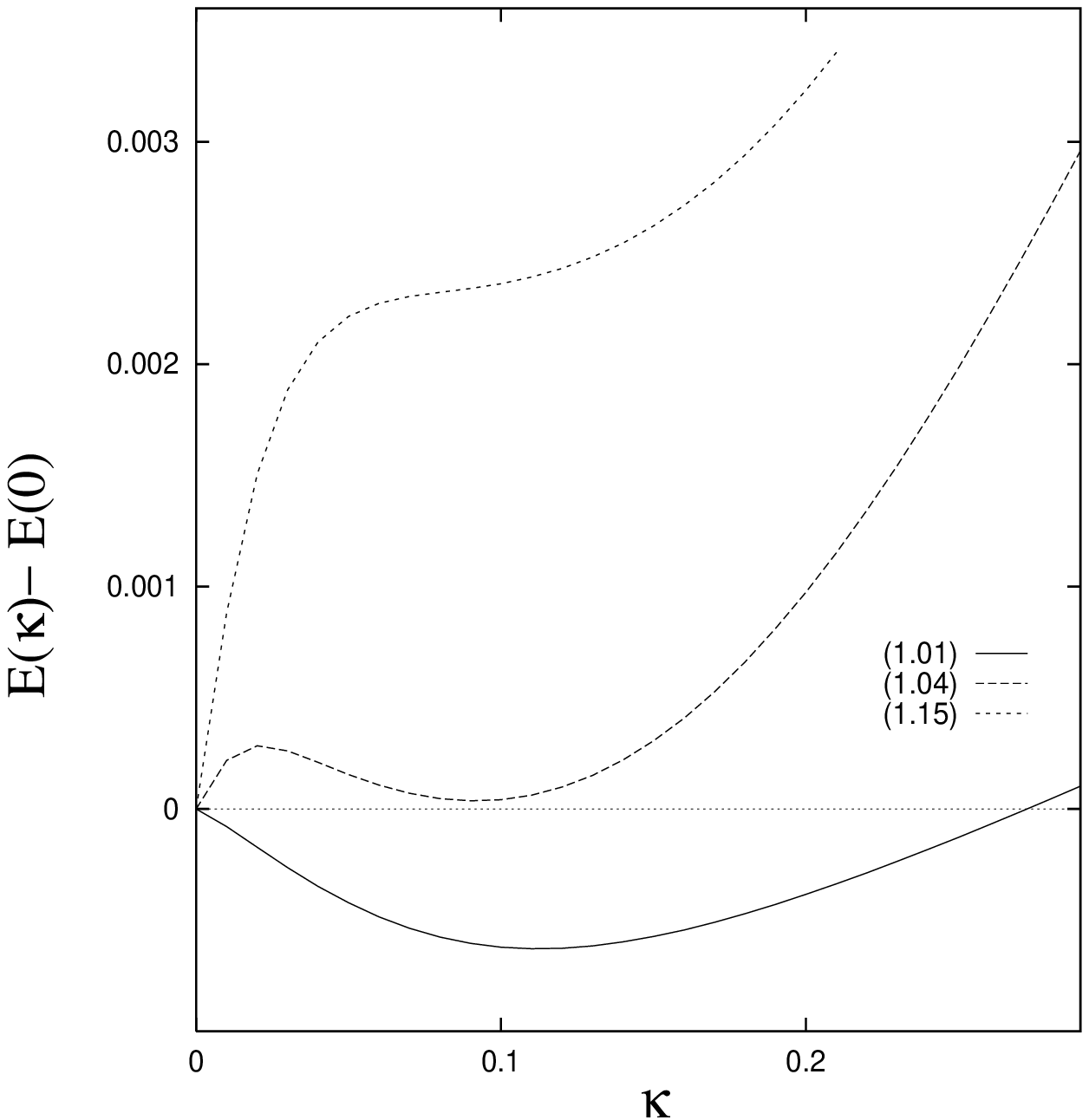}}{\vskip0inFig. \ 1~~ The minimised
energy $E(\kappa)$ as a function of $\kappa$ for the triangular lattice at
$\nu=1.15$, $\nu=1.04$ and $\nu=1.01$. The magnetic field $B = 7$ T.}

To highlight the role of the `size' parameter $\kappa$, we refer to Fig.\ 1, a plot
of $E(\kappa)$, the energy of the N\'eel triangular lattice minimized with respect to
$\lambda$, as a function of $\kappa$ for three different values of $\nu$. As $\nu \to
1$, the lattice gets very dilute, and the energy $E(\kappa)$ attains its minimum at a
value of $\kappa$ close to the single skyrmion value. Next consider the dense region
when $\kappa^{-1} > a$ (strongly overlapping skyrmions).  The nonadditivity of spins
in this region, implies that $E_Z$ is insensitive to $\kappa$. The self-part of
$E_{coul}$ and $E_0$ however, increase linearly with increasing $\kappa$ (for small
$\kappa$) as in Fig.\ 1.  Keeping $a$ fixed, a further increase in $\kappa$ will
separate the skyrmions from each other, giving rise to another minimum at the single
skyrmion value of $\kappa$. These minima become degenerate at $\nu \approx 1.05$. A
further decrease in $\nu$ makes the energy of the second minimum lower. This suggests
that there are two different triangular crystals associated with $\kappa = 0$
(``fat'' skyrmions) and $\kappa > 0$ (``thin'' skyrmions). For the square lattice,
the ``thin'' skyrmion solutions have lower energy throughout the $\nu$ range of
physical interest. 

Figure 2 shows the minimised energies of the N\'eel square and the N\'eel triangular
crystals as a function of $\nu$. When $\nu$ is away from $1$, a triangular lattice
with $\kappa=0$ ($\Delta_1$ phase) has the lowest energy. The smaller Zeeman energy
of the competing square lattice with $\kappa > 0$ is offset by $E_0$ and $E_{coul}$.
As $\nu$ approaches $1$, $E_{coul}$ decreases, resulting in a weak first-order
transition (slope discontinuity) to a square lattice with $\kappa > 0$ at $\vert
\nu-1 \vert \approx .05$ (for a $B = 7$ T). Thus the structural transition between
the $\Delta_1$ and the square phase is accompanied by a change in the size of the
individual skyrmions. In the region of parameter space which favours the square
phase, the $\kappa > 0$ triangular lattice loses out on a higher $E_0$ and $E_Z$. As
$\vert \nu-1 \vert$ gets very close to zero, the difference in energy between these
two lattices vanishes.  In the limit of extreme dilution, $\nu \rightarrow 1$, the
energy of the $N$-skyrmion configuration can be evaluated as an expansion in $(\kappa
a)^{-1}$,
\begin{equation} 
E_{N} = N E_1 + e^{*} \sum_{i > j}
\frac{1}{\vert {\bf x}_i - {\bf x}_j \vert} + {\cal O}\left(\frac
{1}{\kappa a}\right)\,\,, 
\label{eq:dilute} 
\end{equation} 
where $E_1$ is the single skyrmion energy and the coulomb interaction is between
point charges. To leading order, $E_N$ is clearly minimised by placing the charges on
a triangular lattice ($\Delta_2$ phase), however it could have any orientational
order. Higher order terms favour a N\'eel orientation. Thus a second structural
transformation should occur between the square N\'eel lattice and a triangular N\'eel
lattice with $\kappa > 0$ ($\Delta_2$ phase) via a weak first order phase transition
as $\nu$ approaches $1$. 
\myfigure{\epsfysize2.5in\epsfbox{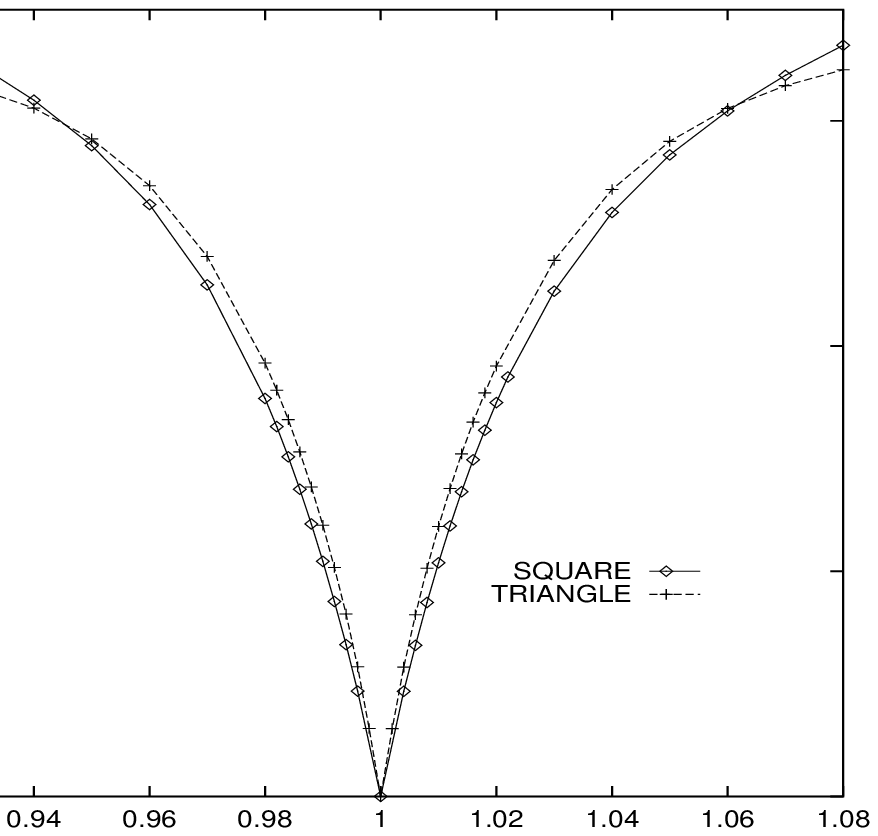}}{\vskip0inFig. 2~~Energy density (in
units of meV/ $l_c^2$) as a function of $\nu$ for the triangular ($\Delta_1$) and
square crystals at $B=7$ T.}

In the vicinity of the $\Delta_1$-square transition, we find that $\kappa^{-1} \sim
a$. A simple scaling argument shows that the shape stiffness of a single skyrmion is
of the same order as the elastic stiffness of the crystal in the neighbourhood of the
transition. The single skyrmion energy is of the form
\begin{equation} 
E_1 = \gamma c_1 + \frac {g^* c_2 R^2}{2}+\frac {e^* c_3}{R}
\end{equation} 
where $R$ sets the scale for the size of the skyrmion and $c_1, c_2$ and $c_3$ are
constants of ${\cal O}(1)$. At ${\bar R} \sim a$, the shape stiffness, $\partial^2
E_1/ \partial R^2\vert_{R=\bar R}$, (where $\bar R$ minimises $E_1$) is comparable to
the elastic stiffness of the crystal $\sim e^2/Ka^3$. The novel feature of this
structural transition is that it is caused by the shape deformability of the
``atoms'', revealing a richer physics than that of Wigner crystallisation. 

A computation of the spin polarization of the skyrmion crystals at $T=0$ (Fig.\ 3),
agrees well with the experimental data of \cite{barret} obtained at $1.5$K.  Since
the spin polarization is a measure of the size $R$ of the individual skyrmions, this
indicates that $R$ does not change significantly over this temperature range. An
interesting feature is that the $\Delta_1$ ($\kappa=0$) to square ($\kappa >0$)
structural transition is accompanied by a discontinuity of about 5-10\% in the spin
polarization, and so may be probed by accurate spin polarization measurements. Since
thermal fluctuations and the presence of quenched disorder smear out this
discontinuity, it may be necessary to go to very low temperatures to see this effect. 
\myfigure{\epsfysize2.5in\epsfbox{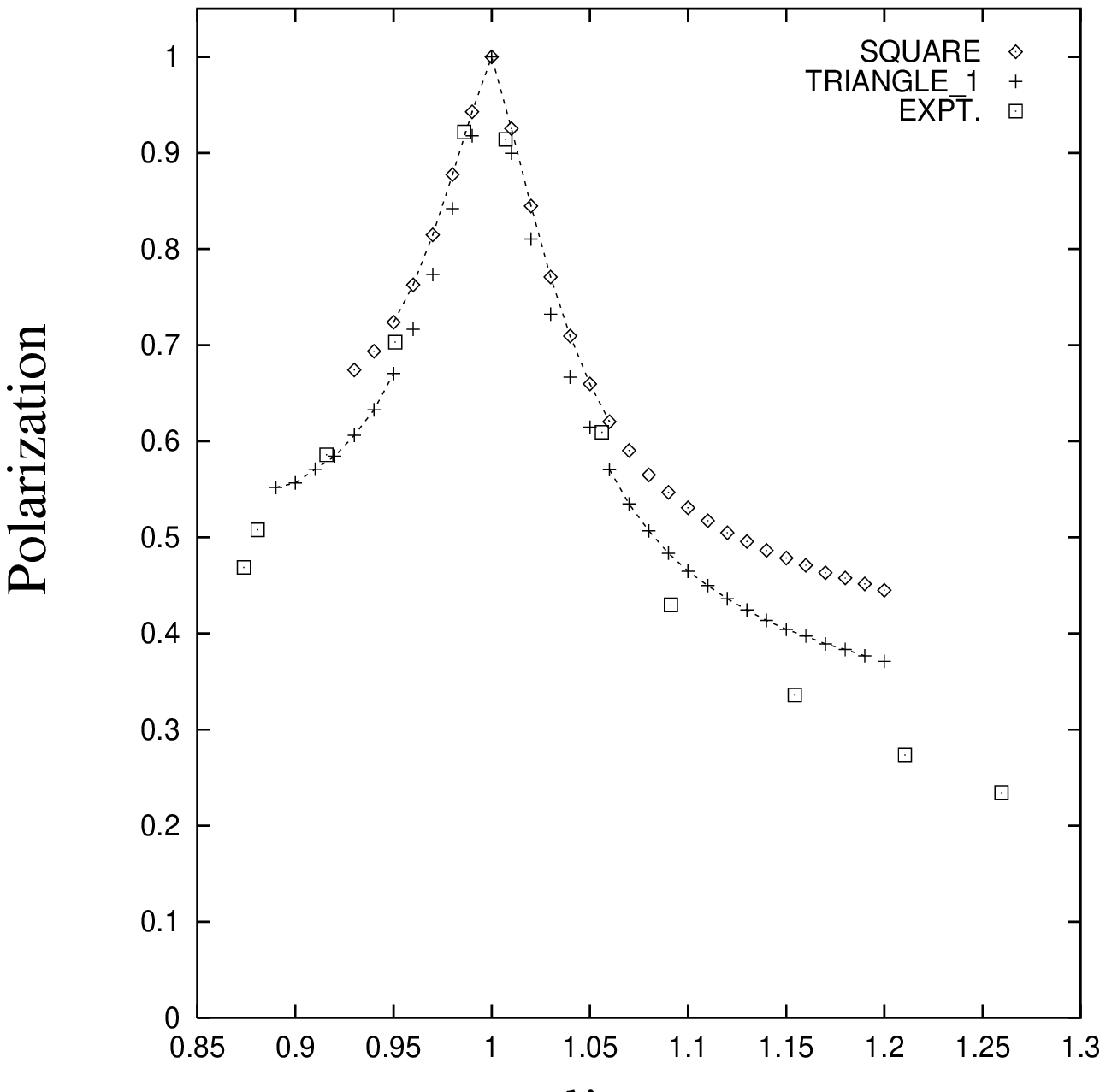}}{\vskip 0inFig.\ 3~ Spin polarization
as a function of $\nu$ at $B=7$ T for the square and triangular lattices compared to
the experimental points taken from Ref.\ \cite{barret}. The structural transition
predicted by our theory is accompanied by a discontinuity in the spin polarization at
$\nu \approx 0.95$, as seen from the dashed curve.}
 
How `good' is our classical variational ansatz Eq.\ \ref{eq:nsky} ? A more general
ansatz for the individual skyrmions would involve deviations of the shape from
circular symmetry, and giving each skyrmion an arbitrary topological charge $Q$. It
is easy to see that both these generalisations lead to an increased energy, in the
region where the skyrmions do not overlap. 

How do quantum and thermal fluctuations affect the classical $T=0$ phase diagram
presented above? The qualitative features of the phase diagram in the $T - \nu$ plane
(Fig.\ 4) may be glimpsed from general arguments. Moving away from $\nu =1$ along the
$T = 0$ axis, reduces the lattice spacing, and shrinks the skyrmion size (due to
coulomb repulsion). At a critical lattice spacing, zero-point fluctuations would melt
the crystal. Quantum melting in a strong magnetic field would ensue when the magnetic
length $l_c \sim \alpha a$, where $\alpha \sim 0.1 - 0.2$. This corresponds to an
energy scale of $(\hbar/0.1a)^2m^{-1}$, where $m$ is the mass of the skyrmion. At the
melting transition, this should compare with the coulomb energy $e^2/a$. We estimate
the quantum melting transition (QM) to occur at $\nu \sim 0.8$. This melting into a
quantum oriented liquid (QL) is most likely continuous with an intermediate quantum
hexatic phase. Beyond this ofcourse, the description in terms of skyrmions breaks
down as $\nu$ approaches the next quantum hall plateau, e.g. $2/3$. In addition, we
encounter a new quantum orientation disorder transition (QOD) in the limit of extreme
dilution. The oriented (N\'eel) crystal is characterised by power-law correlations in
the sublattice orientation. Quantum fluctuations would destroy this order when the
fluctuations in the orientation become of the order of $ 2\pi$. i.e. when $
\hbar^2/2I \sim J(a)$, where $I$ is the moment of inertia of the skyrmion and $J(a)$
is the energy cost in changing the orientation (and is a decreasing function of $a$).
This leads to a quantum disoriented crystal in the dilute limit via a Kosterlitz -
Thouless transition. 
 
An increase in temperature, $T$, would result in an oriented crystal with quasi-long
range order in position and orientation. The advantage that the square phase had over
the $\Delta_2$ phase at $T=0$ now diminishes, since the renormalized $\gamma$ gets
weaker and the coulomb interaction remains relatively unaffected. Thus at higher
temperatures, the square phase disappears, giving rise to an iso-structural
(first-order) phase boundary, where the two triangular N\'eel phases, $\Delta_1$ and
$\Delta_2$ meet. This first-order line terminates on a continuous melting curve (a
Wigner crystal estimate\cite{kogan} gives $T_{m} \sim \sqrt{\vert 1-\nu \vert}$)
which occurs via a defect mediated mechanism (KTNHY\cite{CHAI}). In this case a
hexatic phase (H) intervenes between the solid and the classical liquid (CL) phase.
Indeed there has been a recent suggestion \cite{isos} that the presence of an
iso-structural critical point might reveal a hexatic phase in the vicinity. 
\myfigure{\epsfysize2.0in\epsfbox{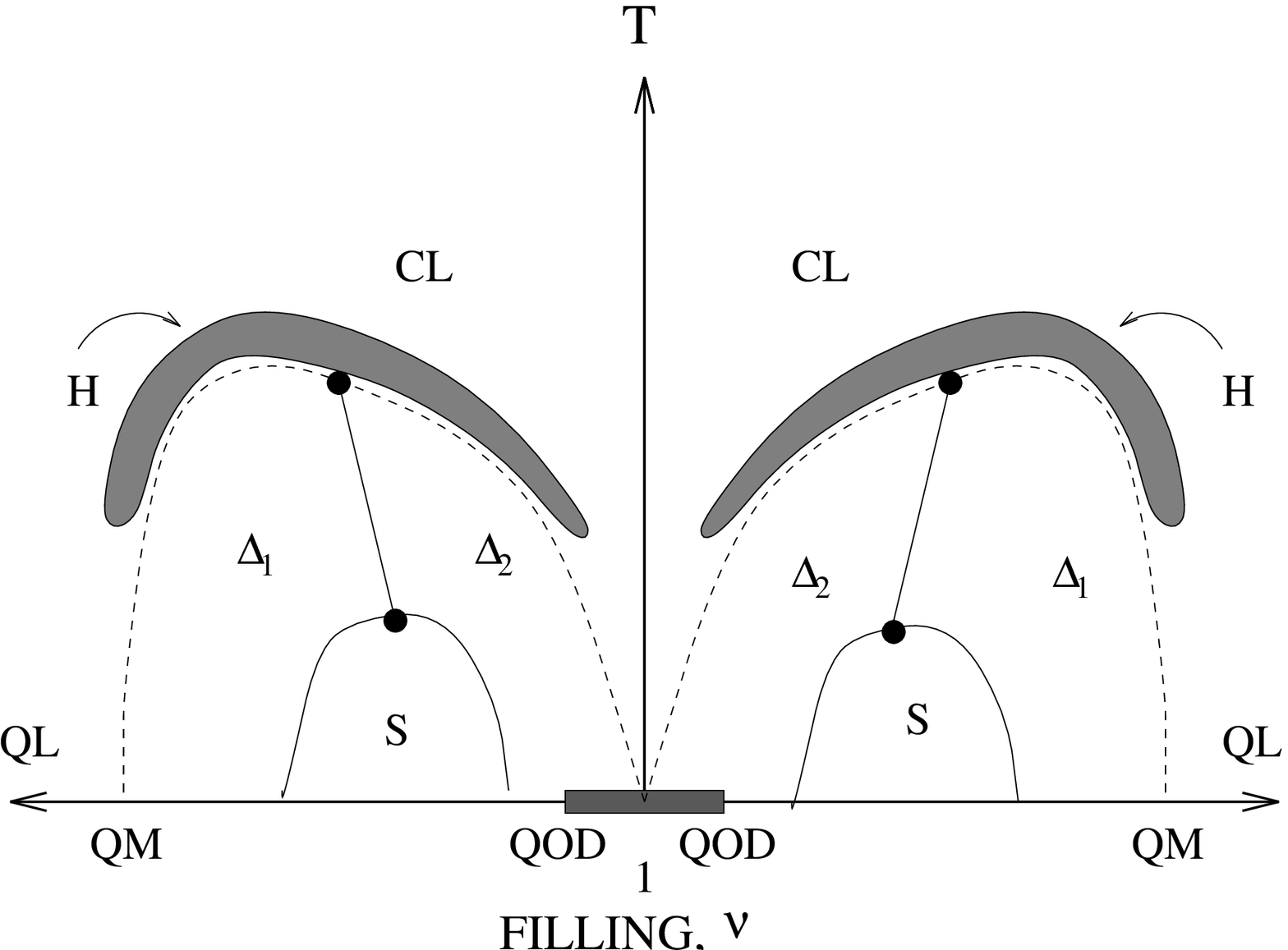}}{\vskip 0inFig.\ 4~~Conjectured Phase
diagram as a function of filling $\nu$ and temperature. The solid and the dashed
lines represent first-order and continuous boundaries respectively. The various
phases shown are defined in the text.}

The application of hydrostatic pressure constitutes an additional `field variable'
\cite{press}. Pressure induces wavefunction overlap which leads to a reduction in
$g$, thus reducing the Zeeman contribution.  We find that when $g \to 0$, the square
N\'eel phase ceases to be a ground state. 

As in the case of the Wigner crystal at $\nu \sim 0.2$ \cite{fert}, we expect that
disorder in the GaAs will not destroy the (quasi)-long range crystalline order of the
skyrmions.  However, the weak first-order structural transitions reported above, will
be rendered continous. We note that the length scales $a$ and $\kappa^{-1}$ are both
around $10^3$ times larger than the GaAs lattice spacing, making the substrate
essentially a continuum. Sliding of the skyrmion crystal would however be prevented
by pinning to the ever present disorder in GaAs. 

We would like to thank B.\ I.\ Halperin for discussions.

\end{document}